\documentclass[baaa]{baaa}

\usepackage[pdftex]{hyperref}
\usepackage{subfigure}
\usepackage{natbib}
\usepackage{helvet,soul}
\usepackage[font=small]{caption}

\contriblanguage{1}

\contribtype{2}

\thematicarea{1}

\title{Signatures of first galaxies at FIR/sub-mm wavelengths}

\titlerunning{First galaxies at FIR/sub-mm wavelengths}

\author{M.E. De Rossi\inst{1,2} \& V. Bromm\inst{3}}
\authorrunning{De Rossi \& Bromm}

\contact{mariaemilia.dr@gmail.com}

\institute{
Universidad de Buenos Aires, Facultad de Ciencias Exactas y Naturales y Ciclo B\'asico Com\'un. Buenos Aires, Argentina
\and CONICET-Universidad de Buenos Aires, Instituto de Astronom\'{\i}a y F\'{\i}sica del Espacio (IAFE). Buenos Aires, Argentina
\and Department of Astronomy, University of Texas at Austin, 2511 Speedway, Austin, TX 78712, USA
}

\resumen{
	Exploramos la posibilidad de detectar las primeras galaxias con un telescopio
	genérico en el infrarrojo lejano y bandas submilimétricas aplicando un modelo
	analítico para la emisión de polvo primordial.  Como se muestra en trabajos
	previos, las galaxias a corrimientos al rojo $z>7$ experimentan una corrección
	K negativa fuerte de forma tal que sistemas de masas similares son más
	brillantes a mayores $z$. Más aún, a una dada masa y $z$, nuestro modelo predice
	que los flujos luminosos crecen proporcionalmente a los cocientes polvo-metal 
	($D/M$) de
	las fuentes primigenias.  Evaluamos la observabilidad de las fuentes del modelo
	a diferentes $z>7$ como función del área de sondeo observada ($\Delta \Omega$) y
	la sensibilidad ($S$) de un instrumento genérico.  Asumiendo $\Delta \Omega \sim 10~{\rm deg}^2$
	y una plausible $S \sim 1~{\mu}{\rm Jy}$ para un relevamiento en el futuro cercano, podríamos
	asegurar la detección de al menos una fuente típica con $D/M \sim 5\times10^{-3}$ a $z>7$.
	Para $S\gtrsim 1~{\mu}{\rm Jy}$ y $\Delta \Omega \lesssim 10~{\rm deg}^2$, 
	valores de $D/M$ mayores que los 
	típicos son requeridos para detectar al menos una fuente individual a $z>7$.
	La observabilidad de las galaxias del modelo es también afectada por la distribución
	de tamaño de los granos de polvo, especialmente a altos $z$.
	}

\abstract{
We explore the possibility of detecting first galaxies with a generic far-infrared/sub-millimeter
telescope by applying an analytical model of primordial dust emission. 
As shown in previous works, galaxies at redshifts $z>7$ experience a strong negative K-correction in such
a way that systems of similar masses are brighter at higher $z$.
In addition, at a given mass and $z$, our model predicts that luminosity fluxes
increase proportionally to the dust-to-metal ratios ($D/M$) of primeval sources. 
We evaluate the observability of model sources
at different $z>7$ as a function of the observed survey area ($\Delta \Omega$)
and sensitivity ($S$) of a generic instrument. Assuming $\Delta \Omega \sim 10~{\rm deg}^2$
and a plausible $S \sim 1~{\mu}{\rm Jy}$ for a near future survey, we could assure the detection of
at least one typical source with $D/M \sim 5\times10^{-3}$ at $z>7$. 
For $S\gtrsim 1~{\mu}{\rm Jy}$ and $\Delta \Omega \lesssim 10~{\rm deg}^2$, 
higher than typical $D/M$ are required to detect at least one individual source at $z>7$.
The observability of model galaxies is also affected by the size distribution of dust grains,
specially towards higher $z$.
}


\begin{document}

\maketitle

\section{Introduction}
\label{sec:Introduction}
With the advent of next-generation observational facilities operating at
different wavelengths (e.g, 
the {\sl James Webb Space Telescope} (JWST), the Square Kilometre Array
(SKA) or the planned {\sl Origins Space Telescope}, among others), the exploration 
of the early Universe is entering a golden age. 
In this context, first stars and galaxies constitute unique tools to 
test galaxy formation models \citep[e.g.][]{bromm2004, bromm2009, bromm2011, dayal2018}.
Thus, it is crucial to develop theoretical models that can make predictions regarding
the nature and properties of primeval luminous sources in order to guide
future surveys. 

Given that {\em typical} dwarf galaxies at $z>7$ are 
expected to have very low dust densities and, thus, low FIR fluxes, 
\citet{derossi2017} showed that their detection will be very challenging with current 
observatories. 
However, primeval galaxies at $z\gtrsim7$ are promising targets for upcoming
observational facilities.
As discussed in \citet[][]{derossi2019}, future space-borne FIR telescopes
could play a crucial role on the exploration of the nature and properties of the first
dust-emitting galaxies at the very dawn of star formation.
In this manuscript, we extend the work by \citet[][]{derossi2019} by analysing in
more detail the dependence of FIR fluxes on the dust-to-metal ratios ($D/M$) of
first galaxies.  We also explore in more detail the observability of model
sources at $z>7$ as a function of the observed sky area ($\Delta \Omega$) and sensitivity 
($S$) of a generic FIR telescope.

\begin{figure*}[!h]
  \centering
  \includegraphics[width=0.40\textwidth]{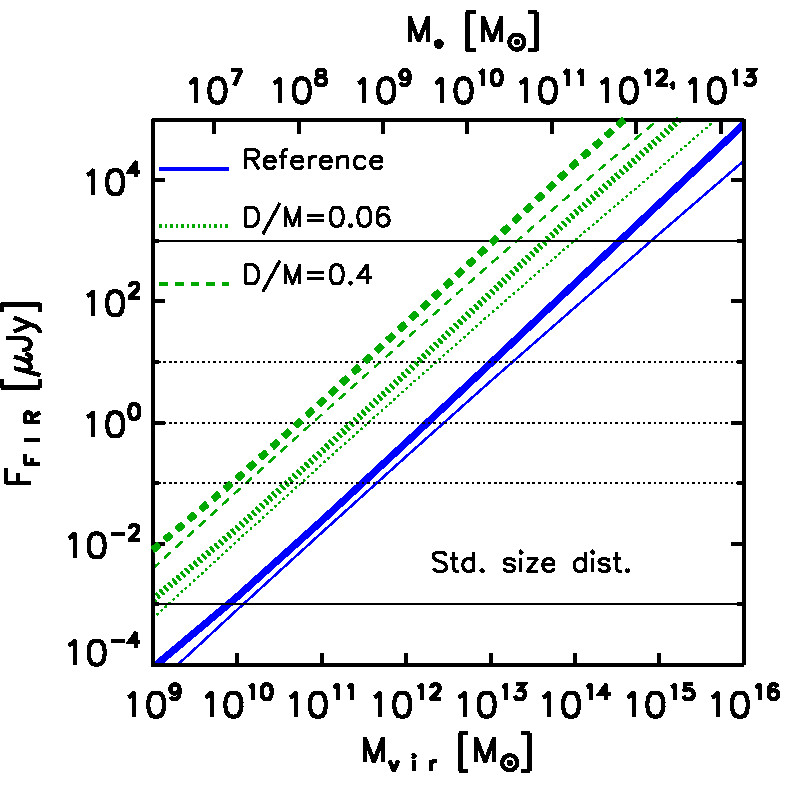}
  \includegraphics[width=0.40\textwidth]{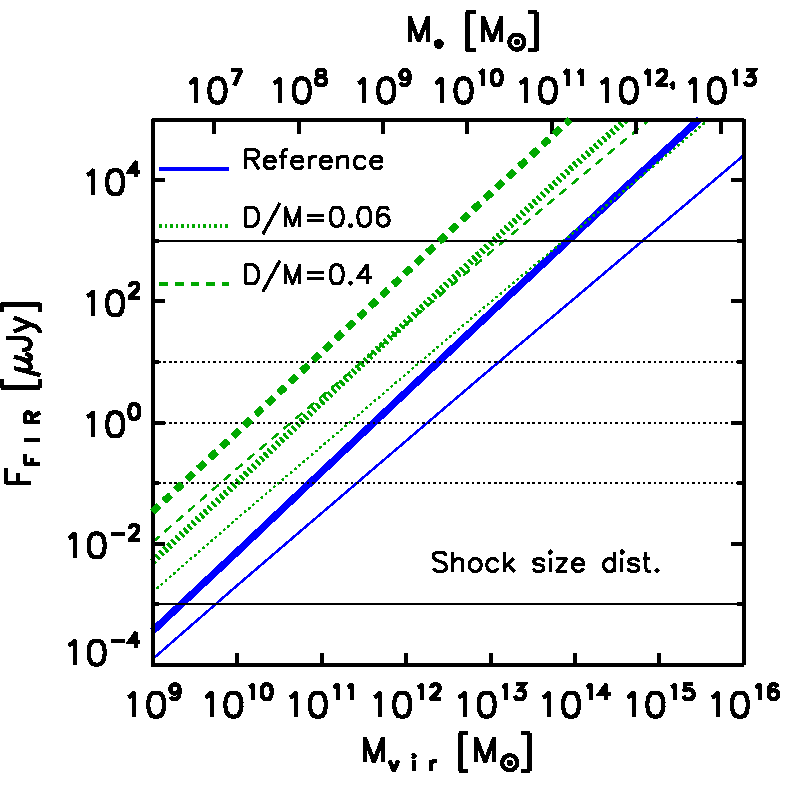}
  \caption{
	  Impact of variations of the dust-to-metal ratio ($D/M$) on the 
	  $F_{\rm FIR}-M_{\rm vir}$ relation for the standard (left panel) and
	  shock (right panel) grain size distributions at $z=7$ (thin lines) and 
	  $z=20$ (thick lines). For comparison, top axes show stellar mass.
	  As a reference, horizontal lines indicate some characteristic 
	  sensitivity values discussed in \citet{derossi2019}.	
}
  \label{fig:Ffir}
\end{figure*}

\section{Dust model}
Dust emission was estimated using the methodology described in \citet{derossi2017} and \citet{derossi2019},
which has proven to be useful for studying high-$z$ galaxies \citep{derossi2018}.
We referred the reader to those papers for a detail description
of the dust model;  here, we only present a brief summary of it.

A model galaxy consists of a dark matter halo hosting a central cluster of Pop~II stars, surrounded by a mixed
phase of gas and dust. 
Our standard model assumes a dust-to-metal mass ratio $D/M = 5 \times 10^{-3}$, a gas metallicity of
$Z_{\rm g}=5\times10^{-3}~{\rm Z}_{\odot}$ and a star formation efficiency of $\eta=0.01$, which are typical values 
expected for first galaxies (\citealt{greif2006, mitchellwynne2015, schneider2016}; see
\citealt{derossi2017, derossi2019} for more details regarding model parameters and the effects of their variations). The spectral energy distribution associated to stars was obtained from {\sc YGGDRASIL} model grids
\citep[][]{zackrisson2011}.
We considered different silicon-based dust models given in \citep[][]{cherchneff2010}. 
However, for the sake of clarity, we only present results corresponding to the
so-called UM-ND-20 model (see \citealt{derossi2017}, for details).
We have checked that other chemical compositions of dust lead to similar general trends.
For the grain-size distribution, we adopted the
‘standard’ and ‘shock’ prescriptions used in \citet{ji2014}.
Dust temperature ($T_{\rm d}$) was determined assuming thermal equilibrium and dust emissivity
was estimated by applying the Kirchhoff’s law for the estimated $T_{\rm d}$ profile.

By comparing plausible sensitivities of a generic instrument with the
average observed fluxes ($F_{\rm FIR}$) of model galaxies at a reference FIR wavelength band 
($\Delta \lambda = 250-750~{\mu}{\rm m}$), 
we determined the lowest virial mass ($M_{\rm vir}$) that a galaxy should have to be detected at a given $z$.
By combining our dust model for individual sources with the Sheth-Tormen mass function \citep{sheth2001}, we
estimated the projected number of detected sources
that are located at redshift higher than $z$, within a given solid angle $\Delta \Omega$ (for a detail description
of this methodology, see \citealt{derossi2019}). 

\begin{figure*}[!h]
  \centering
  \includegraphics[width=0.31\textwidth]{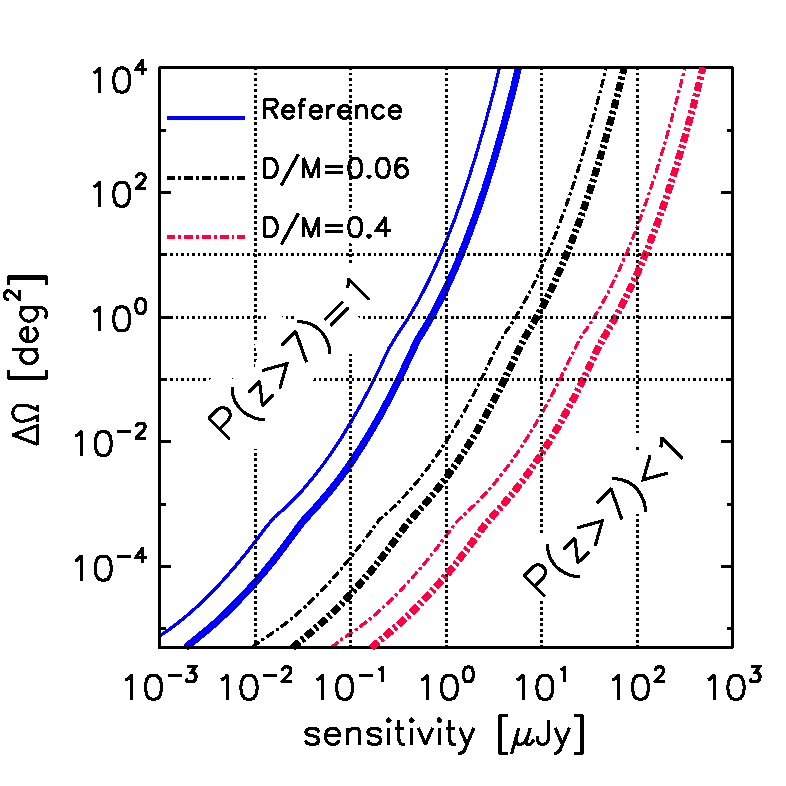}
  \includegraphics[width=0.31\textwidth]{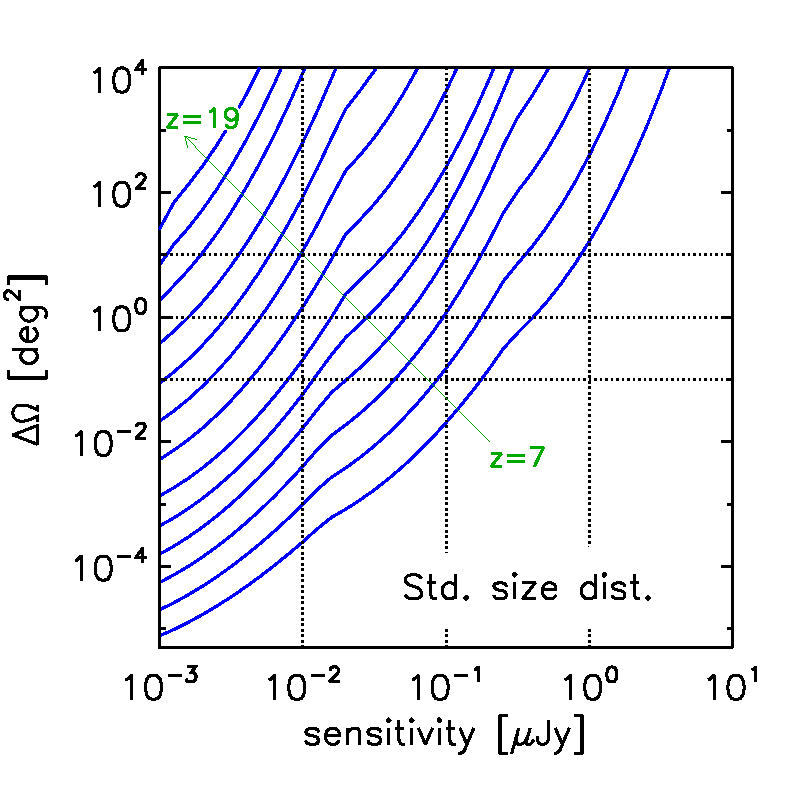}
  \includegraphics[width=0.31\textwidth]{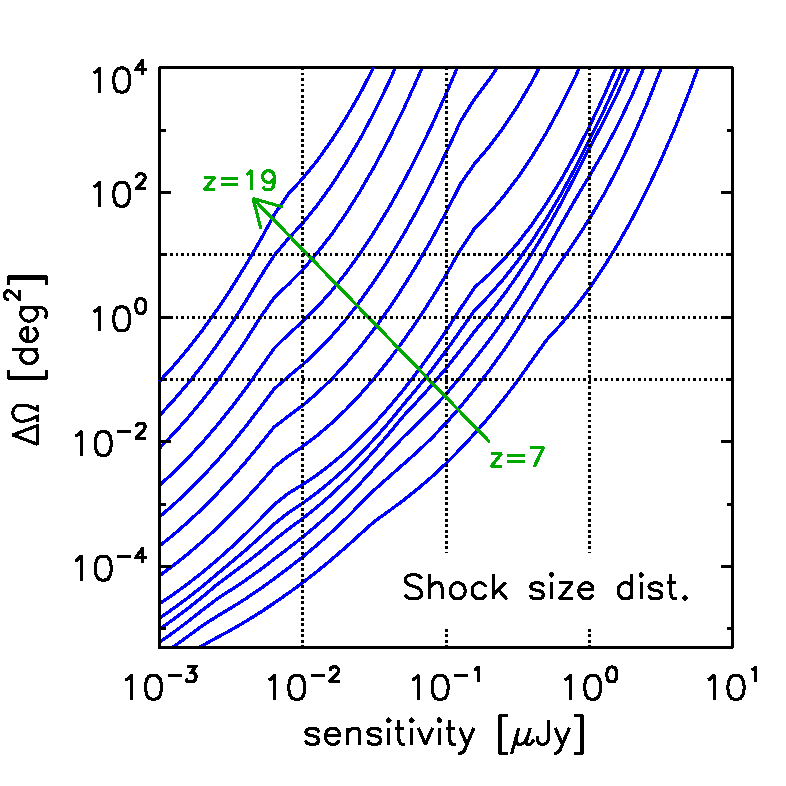}
  \caption{
Constraints on the observability of FIR sources in the $\Delta \Omega$-sensitivity plane.
	Left panel: $\Delta \Omega$-sensitivity curves above (below) which the probability of detecting
	one individual source at $z>7$ is $P(z>7)=1$ ($<1$).  Results are shown for different
dust-to-metal ratios ($D/M$) considering the standard (thin lines) and
shock (thick lines) size distributions.  Middle and right panels:
	analysis of observability of sources at redshifts higher than $z$ for $D/M=5\times10^{-3}$ (reference
	model), 
	considering the standard and shock size distributions, respectively.
	Results are shown for $z=7-19$, with successive curves corresponding to variations
	of $\Delta z$=1 (i.e. $z=7,8, 9...$, etc). 
  }
  \label{fig:omega}
\end{figure*}

\section{Results}
Fig.~\ref{fig:Ffir} shows the $F_{\rm FIR}-M_{\rm vir}$ relation as a function of $D/M$
at $z=7$ (thin lines) and 20 (thick lines). Results for the standard (left panel) and shock
(right panel) grain size distributions are shown.  
We note that, for a given dust model, higher FIR fluxes are obtained for galaxies located at higher $z$. This
is a consequence of the strong negative K-correction affecting primeval dust-emitting source
at $z\gtrsim7$ (see \citealt{derossi2019}, for a discussion).
In addition, at a given mass, FIR fluxes increase
almost proportionally to the increase of $D/M$, with higher fluxes obtained for the shock
size distribution.  Considering that next-generation FIR telescopes could reach a lowest sensitivity of
$\sim 1~{\mu}{\rm Jy}$, we see that a minimum mass of $M_{\rm vir}\sim 10^{12}~{\rm M}_{\sun}$
would be required to detect systems with typical $D/M=5\times10^{-3}$ (reference case).  For 
a higher $D/M=0.4$ (0.06), the minimum mass limit would be 
$M_{\rm vir}\sim 10^{10-11}~{\rm M}_{\sun}$ ($\sim 10^{10-11}~{\rm M}_{\sun}$), with
the exact value depending on $z$ and the grain size distribution.
Such values of $M_{\rm vir}$ correspond to very rare massive galaxies at $z\gtrsim7$, which
might be difficult to be find during blind surveys \citep{derossi2019}.

In Fig.~\ref{fig:omega}, we analyse the observability of primeval galaxies in the
$\Delta \Omega$-$S$ plane.  In the left panel, we show the curves above (below) which
the probability for detecting one individual source at $z>7$ is $P(z>7)=1$ ($<1$), considering
different $D/M$ and grain size distributions.  We can see that, at a given $\Delta \Omega$,
the minimum sensitivity required to assure at least one individual detection increases 
almost proportionally to
$D/M$.  For $\Delta \Omega = 0.1 -10 \ {\rm deg}^2$ and assuming $D/M=5\times10^{-3}$, 
a sensitivity $S\lesssim0.1 - 1~{\mu}{\rm Jy}$ would be required to assure one detection at $z>7$.
For $\Delta \Omega = 0.1 -10~{\rm deg}^2$ and assuming a higher $D/M=0.06$ (0.4),
the required sensitivity would be $S\lesssim1 - 10~{\mu}{\rm Jy}$ ($S\lesssim10 - 100~{\mu}{\rm Jy}$).
In the case of a sensitivity $S\approx1~{\mu}{\rm Jy}$ (which could be a plausible value for a next-generation FIR
telescope) and $D/M= 5\times10^{-3}, 0.06$ and 0.4,
surveys areas $\Delta \Omega \gtrsim 1-10~{\rm deg}^2$,
$\gtrsim 10^{-3} - 10^{-2}~{\rm deg}^2$ and $\gtrsim 10^{-4} - 10^{-3}~{\rm deg}^2$ would
be required to assure the detection of at least one individual source at $z>7$.

In the middle and right panels of Fig.~\ref{fig:omega}, we analyse the observability of sources
located at redshifts higher than a given $z$, for the standard and shock size distributions,
respectively.  A reference value $D/M=5\times10^{-3}$ has been adopted for these panels.
For the standard size distribution, lower sensitivities and larger surveys areas are required 
for detecting sources at all analysed $z$, with such constraints being stronger towards higher $z$.  
Considering a $\Delta \Omega \lesssim 10~{\rm deg}^2$ and a standard (shock) size distribution,
$S \lesssim 1$, 0.1 and 0.01~${\mu}{\rm Jy}$ are required to detect at least one source at
$z>7$, 10 and 14 ($z>7$, 13 and 18).

\section{Conclusions}
We explored the prospects of detecting the dust
continuum emission from first galaxy populations with a generic FIR telescope.
We analyse the FIR observed fluxes of these systems as a function of mass and
redshift, considering also different dust-to-metal ratios and grain size
distributions.  According to our model, FIR radiation from first galaxies
significantly depends on their dust content and dust properties.
Assuming a sensitivity of $S\approx1~{\mu}{\rm Jy}$ for a next-generation telescope,
we obtained that only rare massive systems could be detected at $z>7$.

Combining our dust model for FIR emission of individual galaxies with
the Sheth-Tormen mass function, we evaluated the observability of our galaxy
population in the $\Delta \Omega - S$ plane.  Assuming a sensitivity of 
$S\approx1~{\mu}{\rm Jy}$ and a typical $D/M=5\times10^{-3}$, a 
$\Delta \Omega \gtrsim 10~{\rm deg}^2$ would be required to detect at least
one individual source at $z\gtrsim7$.  However, in the case of higher $D/M$,
smaller survey areas are required.

Our results suggest that future FIR surveys could play a fundamental role on 
constraining the amount of dust in primeval galaxies at $z>7$.

More information and results about this work are available in \citet{derossi2019}.

\begin{acknowledgement}
MEDR thanks the Asociación Argentina de Astronomía for providing
with partial financial support for attending its 61st annual meeting.
MEDR is grateful to PICT-2015-3125 of ANPCyT (Argentina) and also to
Mar\'{\i}a Sanz and Guadalupe Lucia for their help and support.
VB acknowledges support from NSF grant AST-1413501.
We thank Alexander Ji for providing tabulated dust opacities for the different
dust models used here.
This work makes use of the Yggdrasil code \citep{zackrisson2011}, which adopts
Starburst99 SSP models, based on Padova-AGB tracks \citep{leitherer1999, vazquez2005}
for Population~II stars.
\end{acknowledgement}


\bibliographystyle{baaa}
\small
\bibliography{bibliografia}
 
\end{document}